\documentclass[mathleft
]{an}
\usepackage{graphicx}
\usepackage{times}
\usepackage{amsmath}
\usepackage[sort]{natbib}
\bibpunct{(}{)}{;}{a}{}{,}
\begin{document}


\title{On propagation of photons in a  magnetized medium.}

\author{}
\titlerunning{Faraday Effect}
\author{
L. Cruz Rodr\'iguez\inst{1}\fnmsep\thanks{Corresponding author: \email{lcruz@fisica.uh.cu}}
\and
A. P\'erez Mart\'inez\inst{2}\fnmsep\thanks{\email{aurora@icimaf.cu}}
   \and
H. P\'erez Rojas\inst{2}\fnmsep\thanks{\email{hugo@icimaf.cu}}
\and
E. Rodr\'iguez Querts\inst{2,3}\fnmsep\thanks{\email{elizabeth@icimaf.cu}}
}
\institute{Facultad de Fisica,\\
Universidad de la Habana, San L\'azaro y L, Vedado La Habana, 10400, Cuba
\and
Instituto de Cibern\'etica Matem\'atica y F\'isica (ICIMAF) \\
Calle E esq 15 No. 309 Vedado\\
 La Habana, 10400, Cuba
 \and
 Departamento de F\'{\i}sica, CINVESTAV-IPN,\\ Apdo. Postal 14-740, 07000, M\'exico D.F., M\'exico}


\keywords{magnetic fields-Faraday Effect-dispersion relations}

\abstract{%
 The aim of this work is to solve the dispersion relations near the first excitation threshold of photon propagating along the magnetic field  in the strong field limit.  We have calculated the time damping of the photon  in two particular cases: the degenerate gas as well as the diluted gas limit being both important from the Astrophysical point of view. In particular  the diluted gas limit could describe the magnethosphere of neutron stars. The solutions have been used to obtain a finite Quantum Faraday angle in both limits. A resonant behavior for the Faraday angle is also obtained. To reproduce the semi-classical result for the Faraday rotation angle the weak field limit is considered.}

\maketitle

\section{Introduction}

Photon propagating in a magnetized medium  gives rise to the existence of different dispersion relations which  lead to many observable effects.  In the Astrophysical  and Cosmological scenarios one of the most important effects is the Faraday rotation (FR) angle \citep{Giovannini:1997km} which allows to estimate values of the magnetic field in the Universe.

The issue of the light propagation in a material medium like insulators, metals etc has been covered extensively from both the classical and semi-classical point of view. However this approach is not valid anymore when the light travels from inside a compact object characterized by huge densities and strong magnetic field.  An approach from the perspective of  Quantum relativistic theory  should be  included in these studies (\cite{Rojas19821}-\cite{Rodriguez:2013mma} and references therein).

In a previous paper (\cite{Rodriguez:2013mma}) the close relation between Quantum Faraday Effect (FE) and  Quantum Hall Effect  was tackled.
The starting point of that study was the solution of the  Maxwell equation which results from considering the radiative correction: magnetized dense self-energy of photons.  The photon self-energy  gives count of the propagation of the photon through the magnetized medium (electron-positron plasma).

Nevertheless  there are no observations capable of measuring neither the Faraday angle nor the value of the magnetic field  from light that leaves the inner regions of the compact stars. There is also no terrestrial experiments designed to study a similar scenario yet.
However, there are experiments where the FR angle has been measured in Graphene \citep{crassee2011giant}, so it would serve as a test of our results. Hence, our previous paper \cite{Rodriguez:2013mma} had also the purpose to obtain the quantum Faraday angle and QHE in 2D+1 systems (Graphene) from the 3D+1 results.


The aim  of the present paper is to continue  the studies of the propagation of a photon in magnetized medium \citep{Rodriguez:2013mma,ASNA:ASNA201412085} thinking in the Astrophysical implications. In particular we devote this work to  solve the dispersion relation near the first excitation threshold. The  behavior near the thresholds are related to absorption processes and a complete description of the dispersion law of photon requieres to consider them.
For instance, in our first calculation of the Quantum Faraday angle (for a degenerate electron-positron plasma  at strong magnetic field)   \citep{Rodriguez:2013mma} we obtained  a non-finite value of the angle due to the branching points  associated with the absorption processes. As we will  see below these  singularities can be removed finding the solutions near the singular points.
Our study is done in two physical scenarios which are interesting  for Astrophysics:  degenerate gas  ($\mu\gg T$) and diluted gas ($\mu\leq T$). For the diluted gas limit  we consider both: strong ($eB\gg\mu^2$) and weak ($eB\ll\mu^2$) field regime.

The paper is divided as follows.
In Sec \ref{sec2} the solution of the dispersion equation near the first excitation threshold is solved and the Faraday angle obtained in the degenerate and strong field limit is corrected considering the absorption.
Section \ref{sec3} is devoted to compute the scalars $r$ and $t$ in the diluted gas approximation in order to solve the dispersion equation and to find the Faraday angle. The  weak field limit is also studied in the diluted gas approximation with the aim of reproducing the semiclassical results. Finally, in Sec.\ref{sec5} we state the concluding remarks.
\section{Solution of the dispersion equation near the first excitation threshold}
\label{sec2}

The propagation of photons in a relativistic medium at finite temperature and density, in the presence of a constant magnetic field  was studied in \citep{PerezRojas:1979fb,Rojas19821}. The Maxwell equations in the Fourier space are
\begin{equation}\label{Maxwell}
\left[(k^2g_{\mu\nu}-k_{\mu}k_{\nu}+\Pi_{\mu\nu}(k|A_{\mu}^e))\right]a_{\nu}(x)=0, 
\end{equation}
\noindent where $a_{\nu}$ are small corrections to the photon field amplitude, $A_{\mu}^e$ is the external magnetite field and $\Pi_{\mu\nu}$ is the photon
self energy which contains all the information related to the interaction with the medium ($\nu,\mu=1,2,3,4$).

With the aim of solving the dispersion equation in order to find the photon propagation modes, the self energy tensor is diagonalized and a dispersion law for each mode is derived (\cite{Rojas19821}),
\begin{equation}\label{dispersion_final}
k^2=\kappa_{i}(k),
\end{equation}

\noindent where,
$k^2\hspace{-0.1cm}=k_{\parallel}^2+k_{\perp}^2, k_{\perp}^2 \hspace{-0.1cm}= \hspace{-0.1cm}k_1^2+k_2^2, k_{\parallel}^2\hspace{-0.1cm}= \hspace{-0.1cm}k_3^2+k_4^2=k_3^2-\omega^2.$

For the particular case of propagation along the magnetic field $(k_{\perp}=0)$, in the charge asymmetric case, there are three non-vanishing eigenvalues, one
is a pure longitudinal mode and the other two are transverse modes with different dispersion laws, which is the key to the Faraday Effect (FE) (\cite{Rojas19821}).

The eigenvalues related to the transversal modes could be written as $\kappa_{1,2}=t\mp ir=t\pm I_r$, which give rise to the dispersion equation

\begin{equation}\label{dispersioneq}
k_{\parallel}^2=k_3^2-\omega^2=\kappa_{1,2},
\end{equation}
\noindent where,
\[r=iI_r, \quad \hspace{0.2cm} t= -\frac{e^{3}B}{4\pi^{2}}I_{t},\]
\noindent and $I_r$, $I_t$ are the integrals
\begin{equation}\label{intir}
 \hspace{-0.2cm}  I_r=\hspace{-0.1cm}\frac{e^3B\omega}{2\pi^2}\hspace{-0.3cm}\int\limits_{-\infty}^{\infty}\hspace{-0.2cm}dp_3 f\hspace{-0.05cm}(p_3,k_3,B,\omega,\mu)(n_e\hspace{-0.05cm}(\varepsilon_{p,n})-n_p\hspace{-0.05cm}(\varepsilon_{p,n})),
\end{equation}
\[f(p_3,k_{\parallel},\mu,B,T) =\sum_{n,n^{\prime}}F_{n\,n^{\prime}}^{(3)}(0)\left(\frac{k_\parallel^2+2eB(n+n^{\prime})}{D}\right),\]
\[D=[2p_{3}k_{3}+k_{\parallel}^2+2eB(n^\prime-n)]^{2}-4\omega^{2}\varepsilon_{p,n}^{2},\]
\begin{equation}
\hspace{-0.1cm}I_t=\int\limits_{-\infty}^{\infty}dp_3g(p_3,k_{\parallel},\mu,B,T)(n_e(\varepsilon_{p,n})+n_p(\varepsilon_{p,n})), \label{integralIt}
\end{equation}
\vspace{0.1cm}
\noindent
\[g=\hspace{-0.2cm}\sum_{n,n^{\prime}}\frac{F_{n\,n^{\prime}}^{(2)}(0)}{\varepsilon_{n,p}}(1-\frac{(2p_3k_3+J _{nn^{\prime}})(k_\parallel^{2}+2eB(n+n^\prime))}{D},\]
\vspace{0.1cm}
\noindent where $n_{e,p}(\varepsilon_{p,n})\hspace{-0.2cm}=\hspace{-0.2cm}1/\hspace{-0.1cm}(1+e^{(\varepsilon_{p,n}\mp \mu)\beta}),$ are the Fermi-Dirac distribution for electrons and positrons, $n$ is the Landau quantum number, the energy levels are given by $\varepsilon_{p,n}=\sqrt{p_3^2+m^2+2neB}$ and:

$J_{nn^{\prime}}\hspace{-0.1cm}=\hspace{-0.1cm}k_{\parallel}^2+2eB(n^{\prime}-n)$, $F_{n\,n^{\prime}}^{(2,3)}(0)=\delta_{n,n^{\prime}-1} \pm \delta_{n-1,n^{\prime}}.$

\vspace{0.1cm}
Due to the singular denominators in (\ref{intir}) and (\ref{integralIt}), the integral can be divided in a real and imaginary contribution (P\'erez \& Shabad 1982),
 \begin{equation}
  \frac{1}{s-\omega+i\epsilon}=PV\frac{1}{s-\omega}+i\pi\epsilon\delta(s-\omega).
 \end{equation}

 Far from the threshold only the contribution of the principal value of the integral is relevant, but near the singularities the imaginary contribution
 should be considered.

In \cite{Rodriguez:2013mma} an approximate solution of the dispersion equation was derived, regarding the dependence of the scalars $r$ and $t$ with $k_3$ (long wave limit). In particular the strong field ($eB\hspace{-0.1cm}>>\hspace{-0.1cm}\mu^2$) and degenerate ($\mu^2\hspace{-0.1cm}>>\hspace{-0.1cm}T$) limits were considered due to its relevance in astrophysics.
However, only the contribution of the PV of the integral was included. Nevertheless, a complete study of the dispersion equation should include the solution near the threshold.

Close to the branching  points the dispersion equation takes the form:
\begin{equation}\label{disp_threshold}
 k_{\parallel}^2=Im(t)\pm Im(I_r).
\end{equation}
The first excitation threshold which give rise to an infinite behavior is $n^{'}=1$ and $n=0$ or vice versa. Around this branching point the dispersion relation can be written as (\cite{Rojas19821}):
\begin{equation}\label{disp_threshold1}
  k_{\parallel}^2=\frac{e^3B}{4\pi}\frac{k_{\parallel}^{2''}+2eB}{(k_{\parallel}^{2}-k_{\parallel}^{2''})^{1/2}(4m)^{1/2}(m^2+2eB)^{1/4}}N_e^-,
\end{equation}
\noindent where ($k_{\parallel}^{2''}=-[m-(m^2+2eB)^{1/2}]^2$). In the degenerate limit $(\beta\rightarrow \infty)$ the distribution is the difference between two step
functions:
 $$
 N_e^-(\varepsilon_q)=\theta(\mu-\varepsilon_q)-\theta(\mu-\varepsilon_q-\omega).
 $$
After squaring (\ref{disp_threshold1}) we may write it as a cubic equation:

\begin{equation}\label{cubic_equation}
 x^3+px^2+q=0,
 \end{equation}
\noindent where $x=k_{\parallel}^2/m^2$, is a non-dimensional variable and

\begin{equation}
\hspace{-0.4cm} p\hspace{-0.1cm}=\hspace{-0.1cm}-\frac{k_{\parallel}^{2''}}{m^2},\hspace{-0.2cm} \quad q\hspace{-0.1cm}=\hspace{-0.1cm}-\frac{m}{4(m^2+2eB)^{1/2}}\left(\frac{e^3B(k_{\parallel}^{2''}+2eB)}{4\pi m^4}\right)^2\hspace{-0.3cm}.
\end{equation}
Upon the variable substitution $y=x-\frac{p}{3}$ the square term is removed \citep{birkhoff1965survey},
$$ y^3+Ay+B=0,$$
$$A=\frac{1}{3}(3q-p^2),\hspace{0.1cm}B=\frac{1}{27}(2p^3-9pq+27),$$
\noindent and the discriminant is defined as: $ \Delta=A^3/27+B^2/4$, if $\Delta$ is greater than zero the solutions are:
\begin{multline}
  \hspace{-0.3cm} y_1=M+N,\hspace{0.2cm} y_{2,3}=-\frac{1}{2}(M+N)\pm i\frac{\sqrt{3}}{2}(M-N), \\
 \hspace{-0.3cm}  M=\left(-\frac{B}{2}+\sqrt{D}\right)^{1/3},\hspace{0.2cm} N=\left(-\frac{B}{2}-\sqrt{D}\right)^{1/3}. \\
\end{multline}
 Above the threshold the imaginary part of the solution is the responsible for the time damping of the photon $e^{i(\omega t)}$ $\rightarrow$ $e^{i(\omega+i\Gamma)t}$ caused by the excitation of electrons to higher quantum states.

From the equation $\omega+i\Gamma=(k_3^2-x_{2,3}^2m^2)^{1/2}$ we obtain the imaginary contribution to the photon energy (Hugo et al. 1982):
\begin{equation}\label{gamma_eq}
\hspace{-0.2cm} \Gamma =\hspace{-0.1cm}\frac{1}{\sqrt{2}}\hspace{-0.1cm}\left|(m^2Re[x_{2,3}])^2+\hspace{-0.1cm}(m^2Im[x_{2,3}])^2\right|\hspace{-0.1cm}^{\frac{1}{2}}-\hspace{-0.1cm}\left|m^2Re[x_{2,3}]\right|\hspace{-0.1cm}^{\frac{1}{2}}.
\end{equation}
 $\Gamma$ is the probability of absorption of the eingenmodes which remains finite at the threshold and depend only on the value of the magnetic field. In the next section we are going to use this contribution
 in order to continue the previous study of the FR angle (\cite{Rodriguez:2013mma}).
\subsection{Faraday Effect in the degenerate limit}
In the study initiated in \cite{Rodriguez:2013mma} the FR angle,
\begin{equation}\label{faraday_general}
  \theta=\frac{(k_{-}-k_{+})L}{2},
 \end{equation}
 \noindent was obtained from the solutions of the dispersion equation (\ref{dispersioneq}) in the long wave limit ($k_3\rightarrow 0$), regarding the contribution near the branching point. But at this point, using the result derived in the above section, $\Gamma$ can be added to the total photon energy in the expression obtained for the Faraday angle,

\begin{equation}\label{faraday_angle}
  \frac{\theta}{L}\simeq \frac{I_r(\hbar\omega,B)}{2\hbar\omega c}  \quad \rightarrow \frac{\theta}{L}\simeq \frac{I_r(\hbar\omega+i\Gamma,B)}{2\hbar(\omega+i\Gamma) c} .
\end{equation}
This correction was considered for graphene-like systems in \cite{Rodriguez:2013mma}, including it as a phenomenological parameter. But now we are able to estimate the contribution of the absorption process using (\ref{gamma_eq}) for our 3D+1 fermion gas.

\begin{figure}
\includegraphics[width=8.0cm]{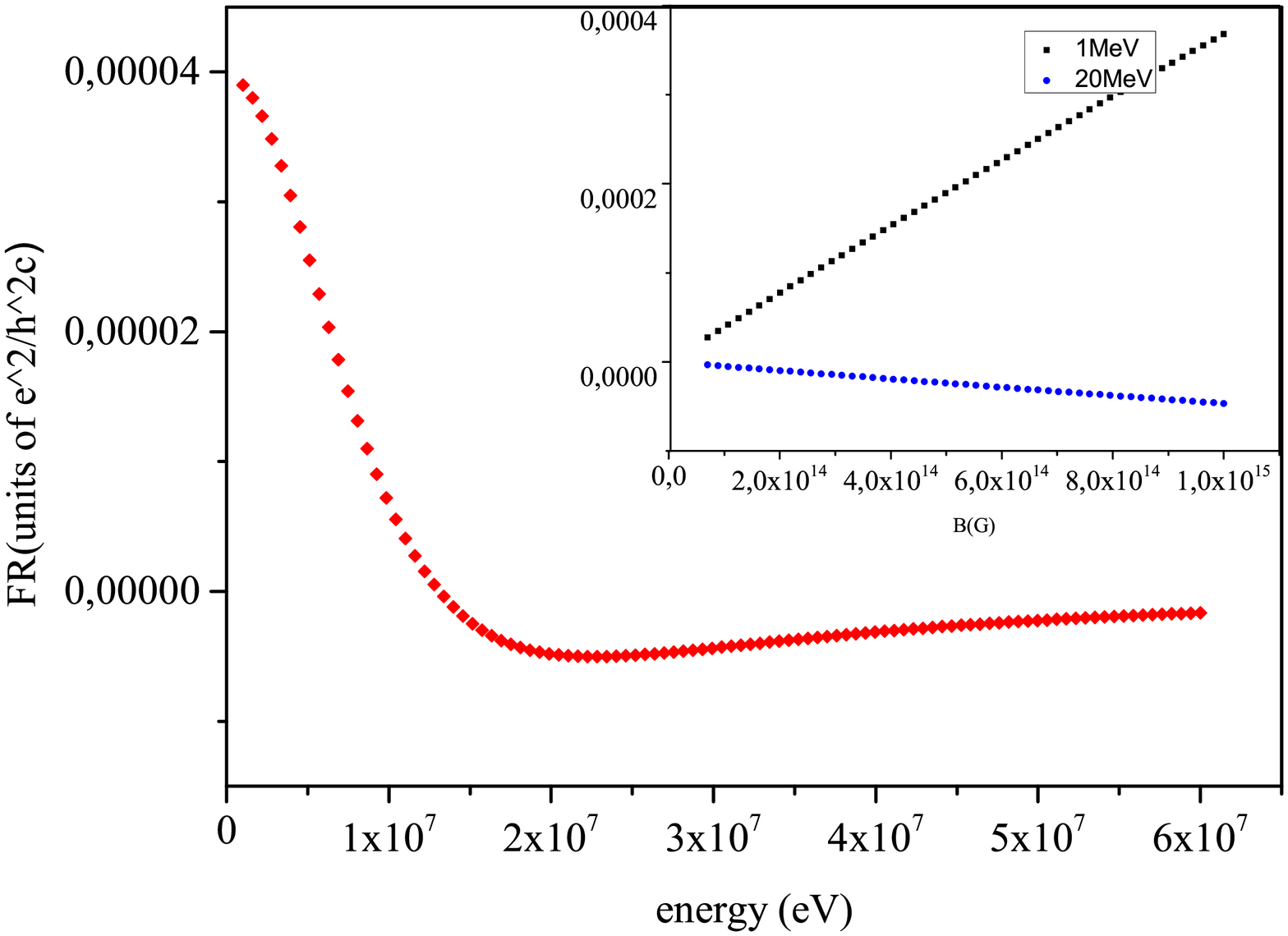}
\caption{(Color on line) FR angle as a function of the photon energy is plotted. Typical values of the magnetic field and  photon energy are considered: $\hbar \omega$ from $1MeV$ to $60 MeV$, $B=10^{14} G$. The value estimated  for $\Gamma$ is $13MeV$.}
\label{figura1}
\end{figure}

 As can be seen in Fig.\ref{figura1} a finite angle is obtained. The enhancement of the Faraday angle is associated with the cyclotron resonance, the curve turn smoothed due to the contribution of the absorption term. The inset figure shows the dependence on \textbf{B} for $1Mev$ and $20MeV$, which is approximately lineal.

This result could be compare to the experimental measurements of FR angle in graphene grown over a substrate \citep{crassee2011giant}
which behaves as a tridimensional system. As we can see Fig.\ref{figura1} shows a good agreement with the behavior of the FR angle reported by \cite{crassee2011giant}.

\section{Dispersion equation in the diluted gas limit.}\label{sec3}
 Another interesting limit from the astrophysical point of view is the photon propagation in a relativistic and diluted media ($T>>\mu$), where strong magnetic field exist. This is the case of radiation propagating through the magnethosphere of neutron stars.

If ($T>>\mu$) the particles distributes classically,
\begin{equation}\label{Maxwel_boltzman}
 n_{e,p}=\frac{1}{1+e^{(\varepsilon_e\mp\mu)\beta}}=e^{-(\varepsilon_e\mp\mu)\beta}.
 \end{equation}
 After a Taylor expansion of the functions $f$ and $g$ in (\ref{intir}) and (\ref{integralIt}), using the fact that the main contribution to the integrals arise from the $p_3=0$ term, so the integrals $I_r$ and $I_t$ can be written as:
\begin{equation}\label{auxiliar1}
I_r=\frac{eB\omega}{2\pi^2}f(0,k_3,\omega,\mu,B,T)\int dp_3  (n_{e}-n_{p}),
\end{equation}
\begin{equation}\label{auxiliar2}
I_t=g(0,k_3,\omega,\mu,B,T)\int dp_3 (n_{e}+n_{p}).
\end{equation}

\noindent By expressing the integral in terms of the modified Bessel function, we get
\begin{equation}
      \int_{-\infty}^{\infty} dp_3  n_{e,p}\label{auxiliar}=\frac{4m_ne^{\pm\beta\mu}}{\sqrt{\pi}}K_1(m_n\beta)\Gamma(3/2),
\end{equation}
\noindent where 
\begin{equation}\label{Bessel}
  K_{\nu}(z)=\frac{\sqrt{\pi}z^{\nu}}{2^{\nu}\Gamma(\nu+1/2)}\int_1^{\infty}e^{-zt}(t^2-1)^{\nu-1/2}dt,
\end{equation}
\noindent and $m_n=\sqrt{m^2+2neB}$.

In the strong field limit only the contribution of the lowest Landau level (LLL) remains, which means to consider in (\ref{intir}) and (\ref{integralIt}),
$n=0$, $n^{'}=1$ or vice versa, depending on the Kronecker delta expressions.
Finally in the diluted and strong field limit,
\begin{equation}\label{funcionIr}
I_r=\frac{8e^3B}{\pi^{5/2}}\frac{(k_{\parallel}^2+2eB)\omega m K_1(m\beta)\Gamma(3/2)\sinh(\mu \beta)}{(k_{\parallel}^2+2eB)^2-4\omega^2m^2},
\end{equation}

\begin{equation}\label{funcionIt}
t=\frac{8e^3B}{\pi^{5/2}}\frac{2w^2m^2K_1(m\beta)\Gamma(3/2)\cosh(\mu \beta)}{(k_{\parallel}^2+2eB)^2-4\omega^2m^2},
\end{equation}

\noindent and from (\ref{dispersioneq}), the solution of the dispersion equation is found,
 \[
  \hspace{-4.1cm}k_\parallel^2=\frac{8e^3B}{\pi^{5/2}}K_1(m\beta)\Gamma(3/2)
  \]
\begin{equation}\label{disp_diluted}
 \hspace{0.6cm}\times \left(\frac{2\omega^2m^{2}\cosh(\mu\beta)\pm m\omega(k_\parallel^{2}+2eB)\sinh(\mu\beta)}{(k_{\parallel}^2+2eB)^2-4m^{2}\omega^2}\right).
 \end{equation}
 Holding in mind the condition $(T>>\mu)$,
 \[\sinh(\mu\beta)\simeq \cosh(\mu\beta)\simeq \frac{e^{\mu\beta}}{2},\]
\noindent and as the Bessel function in (\ref{disp_diluted}) converges quickly to zero, the dispersion equation becomes quadratic for each mode, which solution takes the form:
\begin{equation}\label{disp_finaldiluido}
  k_\parallel^2\hspace{-0.1cm}=\hspace{-0.1cm}-\frac{2eB\pm2m\omega}{2}+\sqrt{\hspace{-0.1cm}(2eB\pm2m\omega)^2-2A(B,T,\mu)m\omega},
 \end{equation}
 \noindent where $A(B,T,\mu)=\frac{8e^3B}{\pi^{5/2}}K_1(m\beta)\Gamma(3/2)e^{\mu\beta}$.

 Then, following the same procedure as in the above section, the solution near the threshold is studied. Taking as a starting point the dispersion equation near the first threshold (\ref{disp_threshold1}), but with the proper distribution in the diluted approximation, the value of $\Gamma$ is estimated.

 In the next section the Faraday angle in the diluted gas is studied considering also the solution near the first excitation threshold.

 \subsection{Faraday angle in the diluted gas limit.}
 From the general definition of the Faraday rotation angle (\ref{faraday_general}), where $k_-$ and $k_{+}$ are the wave vectors of each mode obtained from (\ref{disp_finaldiluido}), the following result is derived (units are recovered):
 \begin{equation}
  \frac{\theta}{L}= -\frac{g(eB\hbar c)\hbar\omega}{8((eB\hbar c)^2-(\hbar\omega)^2)},\label{faradaydiluido}
 \end{equation}
$$g(eB\hbar c)=4\frac{e^2}{h^2c^2}\frac{eB\hbar c}{m^2}K_1(m\beta)e^{\mu\beta}.$$

In the diluted limit the Faraday rotation depends on the photon energy, the magnetic field, the particle density and the temperature. The singularities in the denominator are solved adding the contribution of absorption calculated in the previous section.

 The parameters used to evaluate the Faraday angle and the absorption coefficient, correspond  to typical values of magnetic field in the magnethosphere of neutron stars ($B=10^{13}G$) and with radiation emitted in the $\gamma$ region of the spectrum, where the denominator in (\ref{faradaydiluido}) is close to the singularity.

 \begin{equation}
  \frac{\theta}{L}= -Re\left[\frac{g(eB\hbar c)(\hbar\omega+i\Gamma)}{8}\frac{1}{(eB\hbar c)^2-(\hbar\omega+i\Gamma)^2}\right],\label{faradaydiluido1}
 \end{equation}

\begin{figure}
\includegraphics[width=8.0cm]{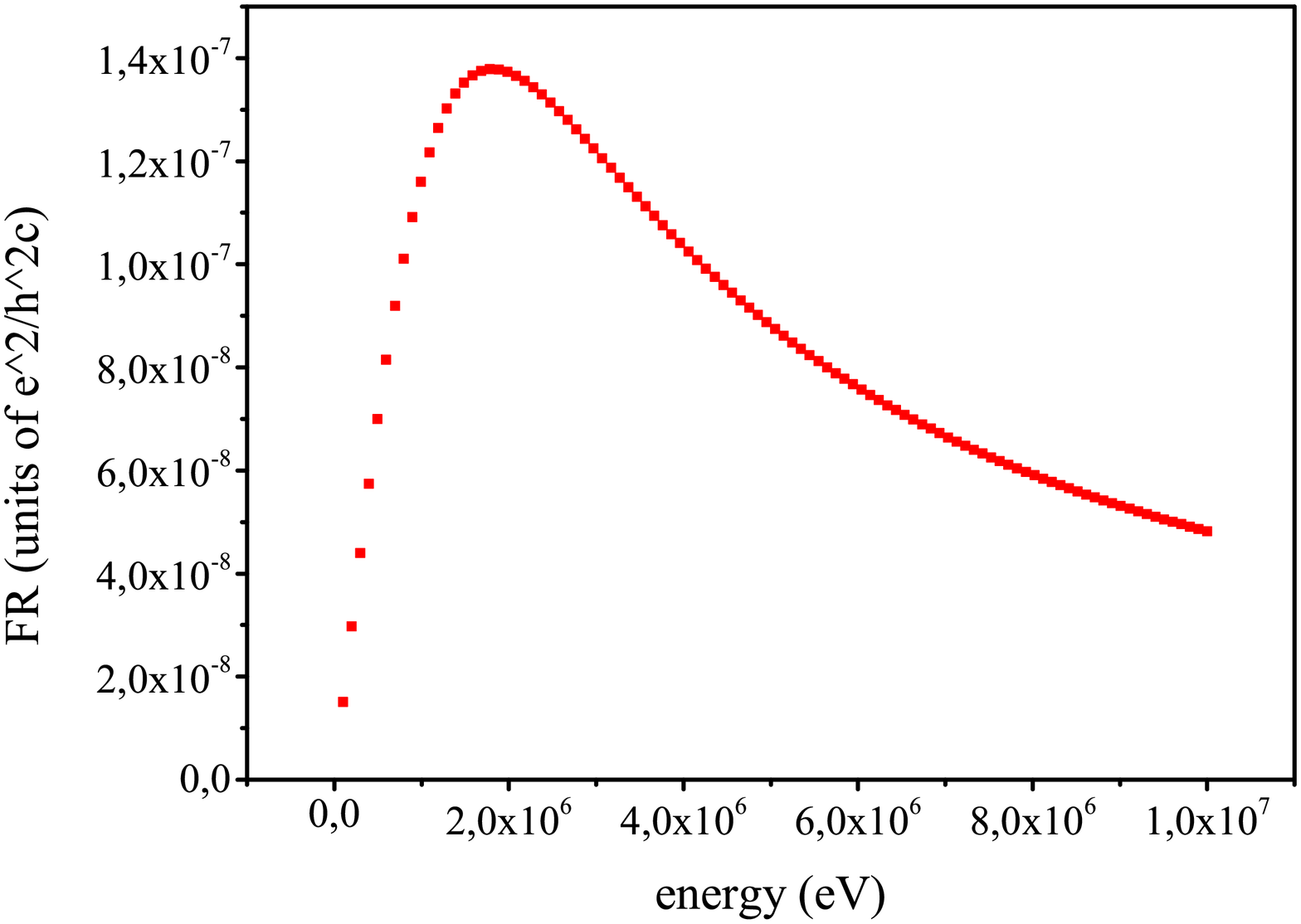}
\caption{(Color on line) FR angle versus photon energy is plotted. The photon energy $\hbar \omega$ runs between $0.1MeV$ to $10 MeV$, the magnetic field $B=10^{13} G$ and $\Gamma=1.8MeV$.}
\label{figura2}
\end{figure}

 In Fig.\ref{figura2} the Faraday angle is plotted as a function of the photon energy in units of $e^2/h^2c$, the maximum value of the angle is close to the region were $\hbar\omega=eB\hbar c$ and it
 is related to electronic transitions to higher Landau levels due to photon absorption.

\subsection{Weak Field Limit}
Starting from (\ref{faraday_angle}) our aim now is to study the weak field and diluted gas limit ($\mu^2<m^2\ll eB\ll T^2$ ) in order to reproduce the semi-classical results for the FE.

 In this limit the energy separation between two consecutive energy levels (quantized by the Landau levels) is proportional to $\sqrt{eB}$ and the spectrum recovers $\varepsilon=\sqrt{p^{2}+m_f^{2}}$,   ($p^2\equiv p_{\parallel}^2+p_{\perp}^2$, $p_{\perp}^2=2|eB|n$) so only the $n=0$ contributes to the sum. The rest of the sum  over Landau levels higher than cero $n\neq 0$  can be replaced by the integration over $p_{\perp}$,
\begin{equation}
   \frac{|eB|}{2\pi}\sum_{n=1}^{\infty}\rightarrow \int\frac{d^2p_{\perp}}{(2\pi)^2}.
\end{equation}

Upon this substitution and considering the classical limit for the particle distribution we have now an integral over the whole momentum space,
\begin{multline}
 I_r = \frac{e^2\omega}{2 \pi^3}\sinh(\mu \beta)\biggl(\int d^3p\frac{J+2p_{\perp}^2}{J^2-4\omega^2(p^2+m^2)}\biggr. \\
 \biggl.-\int d^3p\frac{J^{'}+2p_{\perp}^2}{J^{'2}-4\omega^2(p^2+m^2)}\biggr) e^{-\varepsilon\beta},
\end{multline}
\noindent where $p^2= p_3^2+p_\perp^2$, $J=-\omega^2+2eB$, $J^{'}=-\omega^2-2eB$ and $\varepsilon^2=p^2+m^2$. In the relativistic limit
$\varepsilon^2\simeq p^2$ and the integral may be written in terms of the energy ($d^3p=4p^2dp=4\varepsilon^2d\varepsilon$)
\begin{multline}
 I_r = \frac{2e^2\omega}{\pi^2}\sinh(\mu \beta)\biggl(\int d\varepsilon \frac{(J+\frac{4}{3}\varepsilon^2)\varepsilon^2}{J^2-4\omega^2\varepsilon^2}\biggr.\\
 \biggl.-\int d\varepsilon \frac{(J^{'}+\frac{4}{3}\varepsilon^2)\varepsilon^2}{J^{'2}-4\omega^2\varepsilon^2}\biggr),\label{Ir_campodebil}
\end{multline}
Using the definition of the function $F_n(z)$ (D'Olivo et al. 2003), eq. (\ref{Ir_campodebil}) can be written as
\begin{multline}
 I_r = \frac{2e^2\omega}{\pi^2}\sinh(\mu \beta)\biggl(\frac{4}{\omega^2\beta}\left (J^{'}F_2(z^{'})-JF_2(z)\right)\biggr.\\
     \biggl.  +\frac{4}{3\omega^2\beta^3}\left (F_4(z^{'})-F_4(z)\right)\biggr),
\end{multline}
\noindent where $z=\frac{\beta J}{2 \omega}$, $z^{'}=\frac{\beta J^{'}}{2 \omega}$ and
\begin{equation}
F_n(z)=\int_0^{\infty}\frac{x^ne^{-x}}{x^2-z^2}, \quad E_i(z)=\int_{-\infty}^z \frac{e^tdt}{t},
\end{equation}
\noindent $E_i(z)$ is the exponential integral function.
\[ F_2(z)=1-\frac{z}{2}(e^{-z}E_i(z)-e^zE_i(-z)),\]
\noindent and
\[F_4(z)=2-\frac{z^3}{2}(e^{-z}E_i(z)-e^zE_i(-z)).\]
Expanding around $z=0$, taken into account only the terms of first order in B, using the definition for the Faraday angle given in (\ref{faraday_angle}), the classical limit is derived,

\begin{equation}
 \frac{\theta_F}{L} \simeq\frac{8 e^3B}{\pi^2\omega^2 \beta}\mu,
\end{equation}
\noindent where also the approximation $\sinh(\mu\beta)\simeq\mu\beta $ was used. This is the well known semi-classical result for the Faraday rotation
angle (Ganguly et al. 1999 \& D'Olivo et al. 2003).

\section{Conclusions}
\label{sec5}
We have improved the study of the dispersion relation of photon propagating along the magnetic field. We have obtained the solutions near the first excitation threshold. The study is general but has  been applied in two limit cases: the degenerate gas ($\mu\gg T$) and diluted gas ($\mu\ll T$) due to their relevance  in Astrophysical scenario. Our outcome are the following:
\begin{itemize}
\item The result of the damping time $\Gamma$ related to absorption process  is general (for the first excitation threshold) and it can be used in any particular limit.
\item FR angle can be obtained as a finite quantity after introducing  $\Gamma$ in the singular denominators. On the other hand  FR angle continues showing a resonant behavior.
\item The behavior of the FR angle with the energy in the degenerate limit has a direct analogy with the result reported for Graphene grown over substrate \citep{crassee2011giant}.
\end{itemize}
A future perspective of this work would be to apply our study to  the magnetosphere of the neutron stars described by the diluted gas. The scattering and absorbtion processes of the radiation in that region of the neutron stars  could answer some open question: as cooling, braking index  etc of neutron stars.  Our calculation of the damping time would be useful to develop this task.
\acknowledgements
The authors are very grateful to A. Perez Garcia, J.E Horvath, R.~Pican\c{c}o, H. St\"{o}ecker and J. Rueda for stimulating discussions about the continuation of present work. The work of the authors  have been supported under the grant CB0407 and the ICTP Office of External Activities through NET-35. ERQ has been also supported by a TWAS-CONACYT postdoctoral fellowship.



\end{document}